\documentclass[11pt]{article}
\usepackage{graphicx,amsmath,amsfonts}
\def \beq  {\begin{equation}}
\def \eeq  {\end{equation}}
\def \beqar {\begin{eqnarray}}
\def \eeqar {\end{eqnarray}}
\def\sqr#1#2{{\vcenter{\vbox{\hrule height.#2pt
\hbox{\vrule width.#2pt height#1pt \kern#1pt
\vrule width.#2pt}\hrule height.#2pt}}}}

\def\la {{\langle}}
\def\ra {{\rangle}}
\def\vx {{\vec x}}
\def\vy {{\vec y}}

\def\Tr {{\rm Tr}}

\def\bp {\bar p}

\def\bA {\bar{A}}

\def\bx {\bar{x}}
\def\by {\bar{y}}
\def\bu {\bar{u}}
\def\bv {\bar{v}}
\def\bw {\bar{w}}

\def\vx {{\vec x}}
\def\vz {\vec{z}}
\def\vy{\vec{y}}
\def\vv {\vec{v}}
\def\vu {\vec{u}}
\def\vw {\vec{w}}

\def\del {\partial}
\def\bdel{\bar{\partial}}
\def\a {\alpha}
\def\b {\beta}
\def\e {\epsilon}
\def\d {\delta}
\def\s {\sigma}
\def\l {\lambda}

\def\D {{\cal D}}
\def\bz {{\bar{z}}}

\def\A {{\cal A}}

\def\H {{\cal H}}

\def\G {{\cal G}}

\def\bV{{\bar V}}

\def \H {{\cal H}}

\def\half{\textstyle{1\over 2}}

\begin{document}
\fontfamily{cmr}\fontsize{11pt}{15.6pt}\selectfont
\def \CMP {{ Commun. Math. Phys.}}
\def \PRL {{ Phys. Rev. Lett.}}
\def \PL {{Phys. Lett.}}
\def \NPBProc {{ Nucl. Phys. B (Proc. Suppl.)}}
\def \NP {{ Nucl. Phys.}}
\def \RMP {{ Rev. Mod. Phys.}}
\def \JGP {{ J. Geom. Phys.}}
\def \CQG {{ Class. Quant. Grav.}}
\def \MPL {{Mod. Phys. Lett.}}
\def \IJMP {{ Int. J. Mod. Phys.}}
\def \JHEP {{ JHEP}}
\def \PR {{Phys. Rev.}}
\def \JMP {{J. Math. Phys.}}
\def \GRG{{Gen. Rel. Grav.}}
\begin{titlepage}
\null\vspace{-62pt} \pagestyle{empty}
\begin{center}
\rightline{CCNY-HEP-07/4}
\rightline{April 2007}
\vspace{.6truein} {\Large\bfseries
Yang-Mills Theory in 2+1 Dimensions: Coupling of Matter}\\
\vskip .1in
{\Large\bfseries  Fields and String-breaking Effects}\\
\vspace{.6in}
{\large ABHISHEK AGARWAL$^a$, DIMITRA KARABALI$^b$} and {\large V.P. NAIR$^a$}\\
\vskip .2in
{\itshape $^a$Physics Department\\
City College of the CUNY\\
New York, NY 10031}\\
\vskip .1in
{\itshape $^b$Department of Physics and Astronomy\\
Lehman College of the CUNY\\
Bronx, NY 10468}\\
\vskip .1in
\begin{tabular}{r l}
E-mail:&{\fontfamily{cmtt}\fontsize{11pt}{15pt}\selectfont abhishek@sci.ccny.cuny.edu}\\
&{\fontfamily{cmtt}\fontsize{11pt}{15pt}\selectfont dimitra.karabali@lehman.cuny.edu}\\
&{\fontfamily{cmtt}\fontsize{11pt}{15pt}\selectfont vpn@sci.ccny.cuny.edu}
\end{tabular}

\fontfamily{cmr}\fontsize{11pt}{15pt}\selectfont
\vspace{.8in}
\centerline{\large\bf Abstract}
\end{center}
We explore further the Hamiltonian formulation of Yang-Mills theory in 2+1 dimensions in terms of gauge-invariant matrix variables. Coupling to scalar matter fields is discussed in terms of gauge-invariant fields. We analyze how the screening of adjoint (and other screenable) representations can arise in this formalism.
A Schr\"odinger equation is then derived for the gluelump states which are the daughter states when an adjoint string breaks. A variational solution of this Schr\"odinger equation
leads to an analytic estimate of the string-breaking energy which is within $8.8\%$
of the latest lattice estimates.
\end{titlepage}
\pagestyle{plain} \setcounter{page}{2}

\section{Introduction}

Yang-Mills gauge theories in two spatial dimensions, especially
their nonperturbative properties, are interesting for many reasons.
First of all, they can be a model for the more realistic, but also more 
complicated,
$(3+1)$-dimensional theories. They have nontrivial dynamical content and 
propagating degrees of freedom in contrast to Yang-Mills theories in 
$1+1$ dimensions, yet they are more amenable to mathematical 
analysis with some of the more recent techniques than their 
$(3+1)$-dimensional counterparts. Beyond being a testing ground
for nonperturbative techniques, they are also of direct relevance in,
at least, one physical context, namely, Chromodynamics at high temperatures.
(Recall that in an imaginary-time formalism, all Matsubara modes are 
suppressed at high temperatures, except for the zero modes of bosonic 
fields, leading to a dimensionally reduced theory \cite{high-t}.)
Yang-Mills theories in three (or $2+1$) dimensions are also of interest 
from the point of view of gauge-gravity duality, particularly in the context 
of the recent proposal for string and gravity duals of the maximally 
supersymmetric 
Yang-Mills theory in three dimensions \cite{lm}.
Nonperturbative techniques for strong coupling analyses can 
also help elucidate some of the details of the gauge-gravity duality.

A few years ago, a Hamiltonian analysis of pure Yang-Mills theory in
$2+1$ dimensions was developed based on a gauge-invariant matrix
parametrization of the gauge potentials \cite{KN,
KKN1}. This led to the
calculation of a wave functional for the vacuum state of the theory
\cite{KKN2}. (For reviews, and other approaches and related papers, see
\cite{schulz, others, orland}.) The vacuum expectation value of the Wilson loop
operator could then be evaluated using this wave functional and gave
the result
 \beq \la W_R(C) \ra  = \exp [- \sigma_R {\cal A}_C ]
\label{intro1} 
\eeq 
where $R$ refers to the representation for the
Wilson loop, and ${\cal A}_C$ is the area of the loop $C$. The
string tension $\sigma_R$ was given as 
\beq \sigma_R = e^4 (c_A c_R
/4\pi ) \label{intro2} 
\eeq 
where $e$ is the coupling constant, and
$c_R, ~c_A$ denote the quadratic Casimir values for the
representation $R$ and for the adjoint representation, respectively.
The area law (\ref{intro1}) is consistent with confinement of
charged representations. The calculated value of the string tension
is in very good agreement with lattice estimates \cite{teper,
bringoltz}. More recently, motivated by our Hamiltonian analysis,
there has been an attempt to estimate glueball masses \cite{minic}.
We should note however that the vacuum wave function used in \cite{minic}, which is somewhat conjectural, agrees with ours, derived in \cite{KKN2}, only in the high and low energy limits.\footnote { As we emphasize in the Appendix, in carrying out the regularization of the Hamiltonian and the "local" operators on which it acts, the order in which the regularization parameter of the Hamiltonian and the point-splitting for the operator are taken to zero is important in preserving important symmetries such as Lorentz invariance. We expect that this is one of the reasons for the discrepancies between the two wave functionals.} 
An interesting variant of the Hamiltonian analysis is an `anisotropic' formulation
with two different couplings corresponding to the two spatial  directions \cite{orland}.
The theory is solvable in the extreme anisotropic limit; however, the approach to 
the isotropic
limit is not entirely clear.
In any case, from the results quoted earlier, it is clear that it is
worth pursuing and elaborating on our Hamiltonian approach.

The analysis was done in what may be characterized as a continuum
strong coupling analysis. (This is different from the strong coupling analysis
on the lattice.) For Yang-Mills theory in two spatial dimensions,
$e^2$ has the dimensions of mass and
the expansion parameter is $e^2 /k$ or $k/e^2$, where $k$ is a typical momentum scale.
Modes of momenta much smaller than $e^2$ have to be treated nonperturbatively,
and can be analyzed in a $k/e^2$ expansion, while modes of
momenta much larger than $e^2$ can be treated perturbatively in an
$e^2/k$ expansion.
Continuum strong coupling refers to an expansion treating modes of small momenta
as the dominant contributions to any physical quantity.
The nature of the expansion brings up many questions immediately.
For example, the area law (\ref{intro1}) and the string tension (\ref{intro2}) are obtained
for any representation, including representations of zero $N$-ality
(say, for the gauge group $SU(N)$). But these representations can be screened.
Interpreting the Wilson loop as the propagation of two heavy external charges,
we see that a gluon can bind to the external charge producing a color singlet,
if the external charges are in representations of zero $N$-ality.
(The color singlet state produced by binding gluons to the external charge is
generally referred to as a gluelump.)
Thus, when the separation of the external charges becomes large,
the energy in the string of gauge fields connecting these charges
becomes large and
it is energetically favorable to pair-produce gluons which bind to the
external charges, neutralizing their color charge. There is no potential
between the gluelumps, since they are color-neutral,
and so we do not expect an area law for the
Wilson loop for these representations. The process
of the state of the two heavy  charges connected by the string making
a transition to the state of two gluelumps is
called string-breaking \cite{PW, FK}.
In our previous analysis, we did not see string-breaking.
There is no inconsistency, since we were using the continuum
strong coupling expansion, and 
string-breaking would not be seen to the order we calculated,
but we need to
understand how screening can arise as we improve on this
approximation.

Closely related to this is the question of whether the formula
(\ref{intro2}) for the string tension can be exact in any sense. Are
there corrections to the formula even for representations of nonzero
$N$-ality? There are arguments in the literature, based on the $1/N$-expansion, 
that the ratios of string tensions for different representations should
deviate from the ratios of the Casimir invariants \cite{armoni}.
The breaking of the string for a screenable
representation is a transition from a single color-singlet operator
to two color singlet operators. As such, the transition matrix
element is suppressed at large $N$. So, one possibility is that the
formula for the string tension, for the non-screenable
representations, is exact in the large $N$ limit. This question was
numerically analyzed in detail recently by Teper and Bringoltz
\cite{bringoltz}. They find that, while the formula (\ref{intro2}) for string
tension differs from the lattice result by only about $0.88\%$ as
$N\rightarrow \infty$, the deviations are still statistically
significant. It is therefore important to understand and calculate
possible corrections. The approximation of the wave functional for
modes of small momenta has a simple form, a Gaussian function in the
appropriate variables; this simple form was used in evaluating the
vacuum expectation value (\ref{intro1}). While the closeness of 
the agreement between the lattice results and the values given by (\ref{intro2}) 
shows that the Gaussian approximation is a very good first step, 
the wave functional has non-Gaussian terms and, therefore, it 
has the potential for explaining
corrections to the formula for the string tension.
This is an
issue that deserves further analysis.

In this paper, we will address the problem of screening of
representations of zero $N$-ality, deferring the issue of
corrections to the string tension to a later publication.
As mentioned earlier, screening is due to the formation of color singlet
gluelump states, where gluons which are pair-created from
the vacuum bind to the external charges.
Here we will argue for the possibility of such a bound state from the higher
non-Gaussian terms of the wave functional. We shall then construct
the gluelump state, treating the external charge as a heavy scalar
in the adjoint (or some other screenable) representation. A
Schr\"odinger-type eigenvalue equation will be obtained, within
certain approximations explained in the text. We then carry out a
variational estimate of the gluelump energy. This gives us the
estimate for the string-breaking energy which we compare to the
value obtained from lattice simulations \cite{FK}.

For carrying out this analysis, we need to develop a gauge-invariant
Hamiltonian approach for scalar fields coupled to the Yang-Mills
fields. This will be the subject of the next section. This formalism is
also the first step in extending the continuum strong coupling
analysis of the gauge theory to the case with matter fields added,
and  as such, it is of interest beyond understanding the physics of
screening. For instance, it can lead to an analysis of the
supersymmetric theory, once we
 have further extended this to include fermion fields. 
We may note that there has been a recent proposal
about the gravity dual for ($2+1$)-dimensional Yang-Mills theory with sixteen
 supercharges \cite{lm}. 
 The gauge
 theory in question can be regarded as the dimensional reduction of
 $ \mathcal{N} =$ 4 supersymmetric Yang-Mills theory from $R\times
 S^3$ to $R\times S^2$. A prediction for the leading strong
 coupling contribution to the masses of the operators built out of the 
 scalars fields
has also been computed from the string theory side \cite{lm}. 
Since the gauge-invariant
framework that we shall develop in the paper naturally leads to a
 strong coupling expansion for the gauge theory in the continuum,
there is the potential for obtaining similar results
directly in the strong coupling
regime of the gauge theory and for the possible
confirmation of some details
of the gauge-gravity duality.

We also note that three-dimensional Yang-Mills theory with matter fields in the adjoint representation naturally arises at finite temperatures
 in an effective theory of deconfinement \cite{pisarski}.
The ``matter fields" in this case are the Polyakov loops around the imaginary time 
direction. They take values in the gauge group; nevertheless some of our analysis
can be adapted to analyze that theory. We consider this paper again as 
a prelude to such an analysis.

In section 2, we construct the Hamiltonian for the ($2+1$)-dimensional Yang-Mills theory coupled to adjoint scalars in terms of gauge-invariant variables.
In section 3, we give qualitative arguments for how screening could
arise in the Hamiltonian approach. In section 4, we then use the gauge-invariant
framework outlined in section 2 to work out
the action of the Hamiltonian for
gluelump states of very short spatial extension, appropriate to an
extreme strong coupling approximation. This 
gives a lower bound on the string breaking
energy. The extension of this to gluelump states of finite spatial
extent is given in section 5 where a  Schr\"odinger-type eigenvalue
equation is obtained. In section 6, based on this eigenvalue
equation, we obtain an estimate of the string-breaking energy
and compare it with available lattice data.
There is a short summary/discussion section and the
paper concludes with an appendix with some comments
on how regularization has to be carried out 
for various terms in the Hamiltonian.

\section{Scalars coupled to the YM field}

In this section, we shall set up the essentials of coupling a real scalar field to Yang-Mills theory in two spatial dimensions.
We will consider an $SU(N)$ gauge theory and the scalar field will be in the adjoint representation of $SU(N)$, although various formulae can be easily generalized to any representation.

The action for the theory is \beq S = -{1\over 2e^2} \int d^3x ~ \Tr
(F^2 ) ~+~  \frac{1}{2}\int d^3x~ \Tr \left[ (D_0 \phi )^2 -({\vec
D}\phi )^2 - M^2 \phi^2 \right] \label{phi1} \eeq

The corresponding Hamiltonian is given by
 \beqar
 \H & = & \int d^2x~ \left[ {(E^2 +B^2 )\over 2e^2} + {{\dot \phi}^2 + ({\vec D}\phi )^2 + M^2 \phi^2 \over 2} \right]\nonumber \\
 & = & -\frac{1}{2}\int \left[e^2 \frac{\delta ^2}{\delta {A}^a\delta \bar{{A}}^a} + \frac{\delta ^2}{\delta
\phi^a\delta \phi ^a}\right]+ \int \left[ \frac{{{B^a}
B^a} }{2e^2} + {M^2 \phi^a \phi^a \over 2} + 2 (D\phi)^a (\bar{D} \phi)^a \right]
\label{1} \eeqar In writing this expression, we have made the gauge
choice $A_0 =0$. ${A} = \frac{1}{2}({A}_1+ i{A}_2) , ~\bar{A}
=\frac{1}{2} (A_1-iA_2)$ are the two complex spacial components of
the gauge potential. $D, \bar{D}$ are the holomorphic,
antiholomorphic covariant derivatives \beq D\phi = \del \phi + [A,
\phi]~,~~~~~~~~\bar{D}\phi = \bdel \phi + [\bar{A} , \phi] \eeq In
our conventions, \beq {A} =-i{A}^at^a~, ~~~~\phi = \phi ^at^a \eeq
 where the $SU(N)$
generators are normalized such that
\beq
\mbox{Tr}(t^at^b) = \frac{1}{2}\delta ^{ab}~, ~~~~[t^a,t^b] = if^{abc}t^c
\eeq

We would like now to rewrite the Hamiltonian in terms of gauge-invariant variables following the approach used in  \cite{KN,  KKN1, KKN2} for pure Yang-Mills theory. As before, we use the parametrization
\beq
A= - \del M ~M^{-1}, \hskip .2in  \bA = M^{\dagger -1} \bdel M^\dagger
\label{phi4}
\eeq
If the gauge transformations take values in the Lie group
$G$, $M$ is a complex matrix taking values in $G^{\mathbb C}$, the complexification of $G$.
For the case of $G= SU(N)$, which is what we shall consider in this paper, $M \in SU(N)^{\mathbb C} = SL(N, {\mathbb C})$.

Time-independent  gauge
transformations are realized as left rotations on the matrix
${M}$,
\beq {M}(\vec{x}) \rightarrow
g(\vec{x}) {M}(\vec{x})~,  ~~~~         g(\vec{x}) \in SU(N)
\eeq
As a result, the theory can be entirely rewritten in terms of the gauge-invariant variables
\beq
H = M^\dagger M ~~,~~~~~~~~~~    \chi ={M}^\dagger \phi {M}^{\dagger -1}
\label{5}
\eeq

Before we derive the expression of the full Hamiltonian (\ref{1}) in terms of these gauge-invariant variables, it is useful to first review the case of the pure Yang-Mills theory.
The simplification of the Hamiltonian in terms of the variables $M, M^\dagger$ is as follows.
Introduce the right-translation operator $p_a$ for $M$ and the left-translation operator
$\bp_a$ for $M^\dagger$ by
\beqar
[p_a (\vec{x}), M(\vec{y}) ] &=& M(\vec{y}) (-it_a ) \delta^{(2)}(\vec{x}-\vec{y})\nonumber\\
{}[ \bp_a (\vec{x}) , M^\dagger (\vec{y}) ]  &=& (-it_a) M^\dagger (\vec{y}) \delta^{(2)}(\vec{x}-\vec{y})
\label{phi5}
\eeqar
It was shown in \cite{KKN1} that in terms of these, the kinetic energy operator $T$ may be written as
\beqar
T &=&
-\frac{e^2}{2}\int \frac{\delta ^2}{\delta {A}^a\delta \bar{{A}}^a}  ~
= {e^2 \over 2} \int_{u,v} \Pi_{rs} (\vu,\vv)
\bp_r (\vu) p_s (\vv)\label{phi6} \\
\Pi_{rs} (\vu,\vv) &=& \int_x \bar{\G} _{ar} (\vx,\vu) K_{ab}(\vx) \G _{bs} (\vx,\vv) \nonumber
\eeqar
where $K_{ab} = 2 \Tr (t_a H t_b H^{-1})$ is the adjoint representative of
$H$, and
\beqar
\bar{\G} _{ma} (\vx,\vy)  &=& {1\over \pi (x-y)}   \Bigl[ \d _{ma} - e^{-|\vx-\vy|^2/\e} \bigl(
K(x,\by) K^{-1} (y, \by) \bigr) _{ma}\Bigr] \nonumber\\
\G _{ma} (\vx,\vy)  &=&  {1\over \pi (\bx - \by )} \Bigl[ \d _{ma} - e^{-|\vx-\vy|^2/\e} \bigl(
K^{-1}(y,\bx) K (y, \by) \bigr) _{ma}\Bigr]
\label{phi7}
\eeqar
These are the regularized versions of the corresponding Green's functions
\beq
\bar{G} (\vec{x},\vec{y}) = {1 \over {\pi (x-y)}} ~, ~~~~      G (\vec{x},\vec{y}) = {1 \over {\pi (\bar{x}-\bar{y})}}
\eeq
The parameter controlling the regularization, $\e$, acts as a short-distance cut-off.
Expression (\ref{phi6}) is to be used on functionals where the point-separation
of various factors is much larger than $\sqrt {\e }$. (See Appendix for more details on regularization.) 

As discussed in \cite{KKN1} there is an ambiguity in the parametrization (\ref{phi4}) : $(M ,~M^{\dagger})$ and $(M \bV (\bz),~V(z) M^{\dagger})$  give the same gauge potentials $A, ~\bA$, where $\bV,~V$ are, respectively, antiholomorphic and holomorphic in the complex coordinates $\bz = x_1 + ix_2$ and $z= x_1 -i x_2$. As a result, all physical observables in the theory should satisfy the so-called holomorphic invariance
\beq
H \rightarrow V(z) H \bV (\bz)
\eeq
The particular choice of regularization used in (\ref{phi7}) respects this holomorphic invariance.

When the theory is rewritten in terms of gauge-invariant variables, $p_a, \bp_a$ can be realized as translation operators on $H = M^\dagger M$,
\beqar
[p_a (\vx), H(\vy) ] &=& H(\vy) (-it_a ) \delta^{(2)}(\vx-\vy)\nonumber\\
{}[ \bp_a (\vx) , H (\vy) ]  &=& (-it_a) H (\vy) \delta^{(2)}(\vx-\vy)
\label{phi8}
\eeqar

In \cite{KN, KKN1}, we have argued that the wave functions can be taken to be
functionals of the current $J = (c_A /\pi ) \del H H^{-1}$, where
$c_A$ is the quadratic Casimir invariant defined by
$c_A \delta^{ab} = f^{amn} f^{bmn}$.
The action of the operators $p,~\bp$ on the current $J$ is expressed in terms of the commutation relations
\beqar
 [p_s (\vv),~J_a (\vz)] & =& -i {c_A \over \pi} K_{as} (\vz) \del _z \d (\vz-\vv) \nonumber \\
 {}[\bp _r (\vu),~ J_b (\vw)] & =& -i ({\cal{D}} _w) _{br} \d (\vw-\vu)
\label{8}
\eeqar
where ${\cal{D}}_{w~ab}  = {c_A\over \pi}\partial_w \delta_{ab} +if_{abc}J_c (\vw)$.
Using (\ref{phi6}) and (\ref{8}), the action of $T$ on wavefunctions of the form $\Psi(J)$ can be expressed as
\beq
 T_{YM}~\Psi (J)= {{e^2 c_A} \over {2 \pi}} \left[ \int_z \omega^a(\vz){\delta \over \delta J^a(\vz)}~+\int_{z,w}
\Omega^{ab}(\vz,\vw) {\delta \over \delta J^a(\vz) }{\delta \over
\delta J^b(\vw)}\right] \Psi(J) \label{9} \eeq where \beqar
{}\Lambda _{ra} (\vw,\vz) & = & - \left[ \del _z \Pi_{rs} (\vw,\vz)
\right]
~K^{-1}_{sa} (\vz) \nonumber \\
\omega_a(\vz)& = & if_{arm} \Lambda _{rm} (\vu,\vz)  {\big |}_{\vu
\rightarrow \vz}  \nonumber \\
\Omega_{ab}(\vz,\vw)& = & {\cal{D}} _{w~br} \Lambda_{ra} (\vw,\vz)
\label{10}
\eeqar

The above expressions have been computed in \cite{KKN1} where we found that for small $\epsilon$, $T$ can be further simplified as
\beq
T_{ YM} \Psi (J) = m \left[ \int J_a (\vz) {\d \over {\d J_a (\vz)}} + \int \bigl(
{\cal{D}} _w \bar{G}(\vz,\vw) \bigr) _{ab} {\d \over {\d J_a (\vw)}} {\d \over {\d
J_b (\vz)}} \right]  \Psi (J) + {\cal O} (\e)
\label{phi9}
\eeq
where $m = e^2 c_A /2\pi$.

As for the potential energy term, it is trivially checked that we can write
\beq
V_{YM} = \int d^2x~ {B^aB^a \over 2e^2} = {\pi \over m c_A} \int d^2x~ : \bdel J^a \bdel J^a:
\label{phi10}
\eeq
The regularized form of this expression is
\beqar
V_{YM} &=& {\pi \over {m c_A}} \bigl[ \int_{x,y} \s (\vx,\vy;\lambda ) \bdel J_a (\vx) (K(x,\by)
K^{-1} (y,\by))_{ab} \bdel J_b (\vy) - {{c_A {\rm dim} G} \over {\pi^2
\lambda^{2}}} \bigr] \\
\sigma (\vec{x}, \vec{y} ; \lambda) & = & {e^{{-|\vec{x}-\vec{y}|^2 \over \lambda}} \over {\pi \lambda}} \nonumber
\label{phi11}
\eeqar
where $\sigma(\vec{x}, \vec{y} ; \lambda)$ is a regularized $\delta$-function, $\lambda$ is the parameter of regularization, and we should take the limit
where $\sqrt{\lambda} \ll 1/e^2$. The operator $U^{ab}(\vx, \vy) = [K(x,\by)
K^{-1} (y,\by)]^{ab}$ in (\ref{phi11}) is a "holomorphic" Wilson line connecting the two $\bdel J$'s at different points. $U^{ab} (\vx, \vy)$ is such that the regularized expression for $V_{YM}$ satisfies holomorphic invariance. In considering the application of
$T$ on this, and on other expressions in the rest of this paper, 
the correct procedure is to preserve $\lambda \gg \e$ as we take
these quantities to zero. This is what was used in \cite{KKN1, KKN2}; it is also briefly
explained in the appendix.

Since $T$ as defined in (\ref{phi6}) is valid only for separations much larger than $\sqrt{\e}$,
we need to use $\lambda \gg \e$ to define the potential energy term.
Correspondingly, we can define the kinetic term also with a scale
$\lambda$ by the equation \cite{KKN1}
\beqar
T(\lambda )&=& T + { e^2 \over 2} \log ({2\e / \lambda })~ {\cal Q}\nonumber\\
 {\cal Q}&=& \e \int \s (\vu,\vv;\e) K_{rs} (u, \bv)~\Bigl( \bp_r (\vu) -i \bdel J_r(\vu )\Bigr)~ p_s(\vv)
 \label{phi12}
\eeqar
The understanding here is that ${\l} \gg {\e}$; this
operator is to be taken to act on functionals of $J$ with a point-separation
much larger than $\sqrt{\epsilon}$, with $\e \rightarrow 0$ and
$\sqrt{\l} \ll 1/e^2$.

We now turn to the case with the scalar field added.
Evidently the term $\int { {\delta ^2 } \over {\delta A \delta A}}$ does not change and we can still use the expression
(\ref{phi6}). The full kinetic energy operator $T$ is now
\beqar
T & = &-  \frac{1}{2}\int e^2 \frac{\delta ^2}{\delta {A}^a\delta \bar{{A}}^a} - { 1\over {2}} \int  {{\delta ^2} \over {\delta \phi^a \delta \phi ^a}}
 \nonumber \\
&=& {e^2 \over 2} \int \Pi_{rs} (\vu, \vv) \bp_r(\vu) p_s(\vv ) - {1 \over {2}} \int  {{\delta ^2} \over {\delta
\chi^a \delta \chi ^a}}
 \label{13}
\eeqar
where $\chi$ is the gauge-invariant field defined in (\ref{5}). The action of $p,~\bp$ on $\chi$ is given by
\beqar
[ p_a(\vx), \chi_b(\vy) ] & = & 0 \nonumber \\
{} [ \bp_a (\vx) ,  \chi_b (\vy) ] & = & -f_{abc}~ \chi_c (\vy)
~\delta^{(2)}(\vx-\vy) \label{phi15} \eeqar

The gauge-invariant wave function, $\Psi(J, \chi)$, is now a
functional of both gauge-invariant variables $J$ and $\chi$. In
terms of its action on such functionals $T$ can be written as 
\beq 
T= T_{YM} +i m \int_{z,w} \Lambda_{cd}(\vw, \vz) f^{abc}\chi^a(\vw)
{\delta \over \delta \chi^b(\vw)} {\delta \over \delta J^d(\vz)} -
{1 \over 2} \int  \frac{\delta ^2}{\delta \chi^a\delta \chi ^a}
\label{phi17} 
\eeq 
where $T_{YM}$ and $ \Lambda_{cd}(\vw, \vz)$ are given in (\ref{9}), (\ref{10}).

Similarly the full potential energy term can be expressed in terms
of gauge-invariant fields as 
\beqar
V & = & \int \left( {{B^a B^a} \over {2e^2}} + 2 (D\phi)^a (\bar{D} \phi)^a  + {M^2 \phi^a \phi^a\over 2} \right) \nonumber \\
& = & V_{YM} + \int \left( {2 \pi \over c_A} \bdel \chi^a (\D \chi)^a + {M^2 \chi^a \chi^a
\over 2} \right)
\label{16}
\eeqar

For purposes of easy reference, we shall now collect various formulae 
and write the final form of the Hamiltonian for the Yang-Mills field coupled to adjoint scalars as
\beqar
\H &=& {e^2 \over 2} \int \Pi_{rs} (\vu, \vv) \bp_r(\vu) p_s(\vv ) - {1 \over {2}} \int  {{\delta ^2} \over {\delta
\chi^a \delta \chi ^a}} +V_{YM} + \int \left( {2 \pi \over c_A} \bdel \chi^a (\D \chi)^a + {M^2 \chi^a \chi^a
\over 2} \right)\nonumber\\
&=&m \left[ \int J_a (\vz) {\d \over {\d J_a (\vz)}} + \int \bigl(
{\cal{D}} _w \Lambda(\vw,\vz) \bigr) _{ba} {\d \over {\d J_a (\vz)}} {\d \over {\d
J_b (\vw)}} \right]~+~{\pi \over m c_A} \int  : \bdel J^a \bdel J^a:
\nonumber\\
&&\hskip .2in +i m \int_{z,w} \Lambda_{cd}(\vw, \vz) f^{abc}\chi^a(\vw)
{\delta \over \delta \chi^b(\vw)} {\delta \over \delta J^d(\vz)}\nonumber\\
&&\hskip .2in-{1 \over 2} \int  \frac{\delta ^2}{\delta \chi^a\delta \chi ^a} + \int \left( {2 \pi \over c_A} \bdel \chi^a (\D \chi)^a + {M^2 \chi^a \chi^a
\over 2} \right)
\label{16a}
\eeqar
It is evident from our derivation that this formula for the Hamiltonian is trivially generalizable to the case of several
scalar fields.

\section{Screening: a preliminary qualitative analysis}
 
The next logical step is the application
of the gauge-invariant formalism of the previous section to the
gluelump state. But before doing so, it is useful to
go over some qualitative considerations
of how screening of the adjoint and other
representations of zero $N$-ality could arise in our Hamiltonian
approach.  As discussed in \cite{KKN2}, 
the computation of the expectation value of
the Wilson line in pure Yang-Mills theory is of the form 
\beqar
\la W_R (C) \ra &=& \int d\mu ({\A /\G_*}) ~ e^{-S(H)}~W_R (C)\nonumber\\
&=&\int d\mu (H) e^{2c_A S_{wzw}(H)} ~ e^{-S(H)}~W_R (C)
\label{sc1} 
\eeqar
where $d\mu ({\A /\G_*}) = d\mu (H) e^{2c_A S_{wzw}(H)}$ is the
gauge-invariant volume element for Yang-Mills fields and $e^{-S(H)}
$ is given in terms of the vacuum wave function as $e^{-S(H)} =
\Psi^*_0 \Psi_0$. $S_{wzw}(H)$ is the Wess-Zumino-Witten action for
the hermitian field $H$ and $d\mu (H)$ is the Haar measure for $H$
viewed as an element of $SL(N, \mathbb{C}) /SU(N)$. Explicitly, \beq
S_{wzw} (H) = {1 \over {2 \pi}} \int \Tr (\partial H \bar{\partial}
H^{-1}) +{i \over {12 \pi}} \int \epsilon ^{\mu \nu \alpha} \Tr (
H^{-1}
\partial _{\mu} H H^{-1}
\partial _{\nu}H H^{-1} \partial _{\alpha}H)
\label{sc2}
\eeq
The expression for the wave function $\Psi_0$ shows that
$S(H)$ has the form
\beqar
S(H) &=&  {{4 \pi ^2} \over { e^2 {c_A}^2}} \int  \bdel J_a \left[ { 1 \over {\bigl( m
+ \sqrt{m^2
-\nabla^2 } \bigr)}} \right] \bdel J_a \nonumber\\
&&\hskip .4in  -2 f_{abc} \int f^{(3)}(\vx,\vy,\vec{z})
J_a(\vx) J_b(\vy) J_c(\vec{z})
+{\cal O}(J^4)
\label{sc2a}
\eeqar
The function $f^{(3)}(\vx,\vy,\vec{z})$ has been given in \cite{KKN2}.

Equation (\ref{sc1}) shows that expectation values can be considered as averages in a two-dimensional field theory with the action $S(H)$. In evaluating the string tension and obtaining the formula in terms of the Casimir values of the representations, we have basically used the low-momentum limit of the quadratic
terms (terms which are quadratic in the current $J$) in this functional.
These terms correspond to
\beq
S (H) =  {1\over 4g^2} \int d^2x~ F^a_{ij} F^a_{ij}
\label{sc3}
\eeq
where $g^2 = me^2 = e^4 c_A /2\pi$. In this approximation, the expectation value reduces to the average calculated using a Euclidean two-dimensional Yang-Mills
theory with coupling constant $g$. There are many ways to evaluate these averages, one simple way is to go back to the original variables, the $A$'s, and calculate, for example, in a gauge $A_1 =0$.
In this gauge, the commutator term in $F_{ij}$ is absent and we have a Gaussian functional integral.
This leads directly to the formula for the string tension.

As mentioned in the introduction, the values we find are in very good agreement with the lattice analysis \cite{teper,
bringoltz}, although the lattice results are accurate enough to show a statistically significant deviation, even at large $N$, where the agreement is the best.
Also, it has been argued, on the basis of the $1/N$-expansions, that deviations from the Casimir scaling for string tensions of different
representations are possible \cite{armoni}.
Further, we know, on general theoretical grounds, that the Wilson loops in representations of zero $N$-ality can be screened and will not show the area law behavior for large enough loops.
Therefore, we turn to considerations of possible corrections to this result.
The corrections can arise from the higher terms involving three or more powers of the current $J$.
The string tension is defined by the limit of large Wilson loops. Therefore we may think of ``integrating out'' the high momentum modes in the two-dimensional theory defined by $S(H)$ to obtain an effective low-momentum action $S_{eff}(H)$ and then using this to calculate the average of the Wilson loop.
It is then easy to see that there are
two types of corrections possible. The first set would correspond to a corrected quadratic term, due to terms which modify the ``propagator'' for the currents. This is equivalent to a modified coupling
$g_*$ in the effective two-dimensional Yang-Mills theory. Since this is at the level of the action and  we have not yet introduced the Wilson line,  there is no dependence on the representation of the Wilson line; it depends only on
$N$ and numerical factors. Therefore, these corrections can affect the value of $\sigma_R$, but the ratio
$\sigma_R /\sigma_F$ is unaffected.

The second set of corrections involve the $J^3$ and higher vertices with the currents
connected to the Wilson line via propagators. These cannot be reduced to propagator corrections
and hence we can get representation-dependent modifications. In particular, screening effects have to come from such terms. In principle, all the $J^3$ and higher vertices can contribute.
To elucidate the nature of these terms, consider the contributions from the cubic vertex \cite{KKN2}.
Written in terms of the original variables, the cubic term is of the form
$ \Tr ( {\bar D}^{-1} B [B, {\bar D}B] )$ (Since we have phrased the calculation of the string tension in terms of the $A$'s, we use the same variables for this argument.)
Expanding ${\bar D}^{-1}$ in powers of ${\bar A}$, we see that
\beq
{\bar D}^{-1} = {\bar \del}^{-1} - \bdel^{-1} \bA ~\bdel^{-1} +\cdots
\label{sc4}
\eeq
In the $A_1=0$ gauge, the cubic term is thus of the form
\beq
S^{(3)}_{int} \sim \left[ ({\bar \del}^{-1} - \bdel^{-1} \bA \bdel^{-1} +\cdots )~ \del_1 A_2\right]^a
\del_1A_2^b ~\del_1 \bdel A_2^c ~f^{abc}
\label{sc5}
\eeq
Because of the overall Euclidean invariance,  the Wilson line can be visualized as the propagation of two heavy charges. Consider then the
insertion of the interaction term (\ref{sc5}). The contractions of the $A$'s from the
expansion of ${\bar D}^{-1}$ will produce a sequence of terms which is effectively like the iteration of the kernel in a Bethe-Salpeter equation for a bound state.
In other words, we get terms which can be interpreted as the propagation kernel of a possible
bound state of the gluons with the external charge of the Wilson line.
For the nonscreenable representations, it is not meaningful to think of the bound state in isolation since it is not a color singlet. But for the screenable representations, a color singlet bound state can be formed.
The picture which emerges is one where gluons are created from the vacuum and then bind to the external charges to form color singlets. This can lead to breaking of the single string connecting the external charges into two such singlet bound states.

\begin{figure}[!t]
\begin{center}
{\scalebox{.6}{\includegraphics[280,400][300,500]{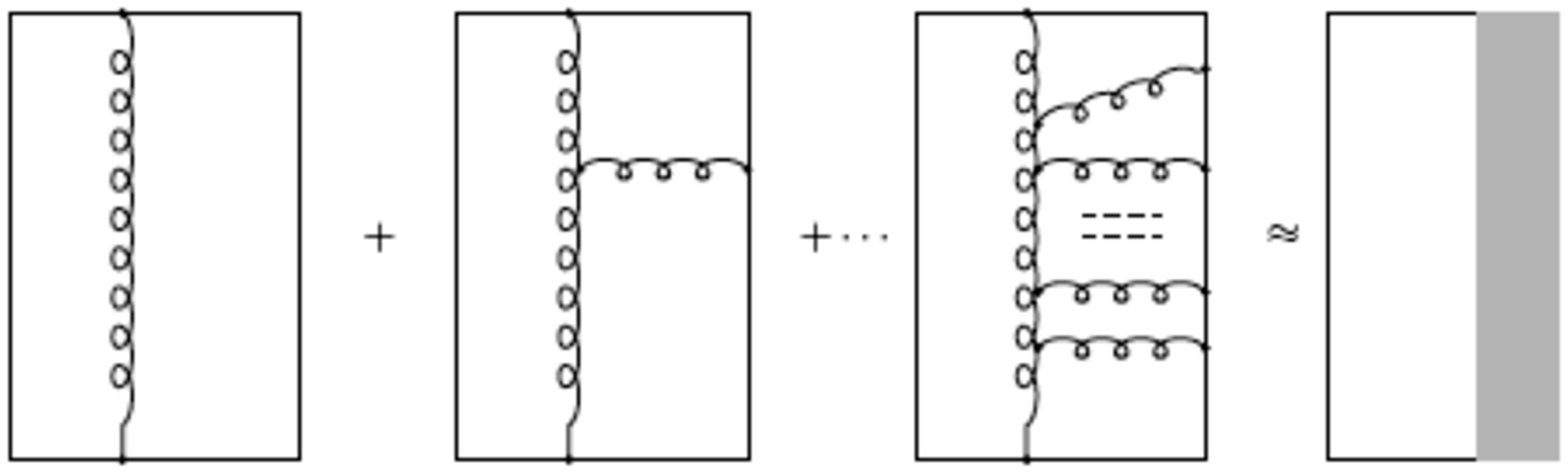}}}\\
\vskip .7in
A set of terms whose sum can be viewed as the propagation kernel of a bound state\\ (shown as thick gray line)
\end{center}
\end{figure}

As for the quantitative calculation of these corrections, we know from the lattice
data that they are small, except for the screening effects.
It is, of course, important to calculate the corrections and provide an analytic justification for why they are small. The calculation of these corrections and the analytic justification for why they are small will be reported in another paper. Here we will pursue the analysis of string breaking for screenable representations.
There are two related issues for the analysis of this question.
There is a transition matrix element between the state of the single string connecting the external charges and the two singlet bound states. This matrix element will describe the process of the string-breaking, and, from general arguments, it should be suppressed by powers of $N$, in a large $N$ expansion. The calculation of this matrix element is quite involved and we will not carry this out here. However, since the energy at which the string breaks corresponds to twice the energy of the singlet bound state, we can calculate this string-breaking energy by calculating the energy of the singlet bound state of a gluon with
the external heavy charge (whose propagation is one side of the Wilson line).
This bound state has generally been referred to as the gluelump. In our language, it corresponds to a wave function of the form
\beqar
\Psi_G &=& \int d^2xd^2y~ f(\vx,\vy)~\bdel J^a(\vx) [K(x, \by ) K^{-1}(y, \by ) ]^{ab} \chi^b(\vy) ~~\Psi_0\nonumber\\
&\equiv& {\cal O}_G ~~\Psi_0
\label{sc6}
\eeqar
Here $J$ represents the gauge-invariant gluon which has a mass $m$.
In reference \cite{KKN1}, we have discussed a constituent picture for the formation of bound states, such as glueballs, in terms of elementary constituents of mass $m$ corresponding to $J$. The gluelump, as described by (\ref{sc6}), is consistent with that picture. The function
$f(\vx, \vy)$ is then the wave function of the constituents in the bound state.

We can now move onto a more specific and quantitative estimate of the bound state energy rather than the general qualitative arguments given above. For this, the strategy will be to
act with the Hamiltonian on the state (\ref{sc6}) and obtain a two-particle Schr\"odinger-like equation for the function $f(\vx,\vy)$ characterizing the bound state.
The details of this analysis are given in the next section.

\section{The action of the Hamiltonian on $\Psi_G$}

We will first consider the (continuum) strong coupling limit where we use the state (\ref{sc6}) with the vacuum wave function $\Psi_0 = \exp (-\half S(H))$ approximated by
\beqar
\Psi_0 &\approx& \exp \left[ - {V_{YM}\over 2m} -{M\over {2}} \int d^2x~ \chi^2\right]\nonumber\\
&=& \exp \left[ - {1\over 8 g^2} \int d^2x~ F^2 -{M\over {2}} \int d^2x~ \chi^2\right]
\label{sc7}
\eeqar
We have used the perturbative vacuum for the $\chi$-field. In the large $M$ limit,
because $M^2 \gg \bdel  \D$, corrections due to the interaction term
in $\bdel \D$ are suppressed by powers of $1/M$.

With the approximation of $\Psi_0$ as in (\ref{sc7}), the action of the Hamiltonian on $\Psi_G$ can be written as\footnote{We take the Hamiltonian to be normally ordered for the scalar fields so that there is no zero-point energy contribution in (\ref{sc7a}).}
\beqar
\H ~\Psi_G &=& M \Psi_G ~+~ {\cal O}_G ~\H_{YM} \Psi_0~+~ \Psi_0 ( T {\cal O}_G )
\nonumber\\
&&\hskip .2in -{e^2 \over 4m} \int \Pi_{rs}(\vu,\vv) [\bp_r (\vu), V_{YM}] [p_s(\vv), {\cal O}_G] \Psi_0 \nonumber\\
&&\hskip .2in  -{e^2 \over 4m} \int \Pi_{rs}(\vu,\vv) [p_s(\vv), V_{YM}]
[\bp_r(\vu), {\cal O}_G] \Psi_0 \label{sc7a} 
\eeqar 

 Since $\Psi_0$ is the vacuum for the Yang-Mills part,
$\H_{YM} \Psi_0 =0$. For the third term and the last term on the
right hand side, it should be kept in mind that $\bp_r$ has
nontrivial action on $\chi$ as given in (\ref{phi15}).

In evaluating $(T {\cal O}_G)$, it is useful to compare it with the action of the Yang-Mills
kinetic term, which we have evaluated elsewhere \cite{KKN1}.
Consider
\beq
\Psi_2 = \int d^2xd^2y~ f(\vx,\vy)~\bdel J^a(\vx) [K(x, \by ) K^{-1}(y, \by ) ]^{ab} \bdel J^b(\vy)
\label{sc7b}
\eeq
Notice that ${\cal O}_G$ is obtained from this by replacing the last $\bdel J$, namely
$\bdel J^b (\vy)$, by $\chi^b (\vy)$.
The action of $T $ on the state (\ref{sc7b}) can be written as the sum of four terms as given below.
 \beq
 T ~\Psi_2 =  I + ~ II + ~ III + ~ IV
 \nonumber
 \eeq
 \beqar
 I &=& {e^2 \over 2} \int \Pi_{rs}(\vu,\vv) [\bp_r(\vu), [p_s(\vv), \bdel J^a(\vx)U^{ab}(\vx, \vy)]]
 ~ \bdel J^b (\vy)~ f(\vx,\vy)\nonumber\\
II&=& {e^2 \over 2} \int \Pi_{rs}(\vu,\vv) [p_s(\vv), \bdel J^a(\vx)U^{ab}(\vx, \vy)]
 ~ [\bp_r(\vu), \bdel J^b (\vy)]~f(\vx,\vy)\label{sc7c}\\
III&=& {e^2 \over 2} \int \Pi_{rs}(\vu,\vv) [\bp_r(\vu), \bdel J^a(\vx)U^{ab}(\vx, \vy)]
 ~ [p_s(\vv), \bdel J^b (\vy)]~f(\vx,\vy)\nonumber\\
IV&=& {e^2 \over 2} \int \Pi_{rs}(\vu,\vv) \bdel J^a(\vx)U^{ab}(\vx, \vy)
 ~ [\bp_r(\vu), [p_s(\vv),\bdel J^b (\vy)]]~f(\vx,\vy)\nonumber
\eeqar
where $U^{ab}(\vx, \vy) =  [K(x, \by ) K^{-1}(y, \by ) ]^{ab}$.
We will consider the evaluation of these terms for small separations
$\vert \vx -\vy\vert \ll 1/e^2$, with $\vert \vx -\vy\vert \gg \sqrt{\epsilon}$. (This means, of course, that the the width of $f$ which controls the average value of $|\vx-\vy|$ is small compared to $1/e^2$.)
Term $I$ then leads to a mass $m$ for $\bdel J$ and the Coulomb potential
${\cal V}(\vx,\vy)$. Terms $II$ and $III$ contain a normal-ordering
term, with the $p, \bp$ acting on the $\bdel J$'s. The other terms arising from
these two are negligible for very small separations; these involve at least one power of
$\vert \vx-\vy\vert$ and three or more $J$'s. Finally the last term gives $m$ for the final
$\bdel J(\vy)$. Thus we find
\beq
 T ~\Psi_2 = \int_{x,y}~ \left[ 2m + {\cal V}(\vx,\vy)\right]f(\vx,\vy)~:\bdel J^a(\vx) U^{ab}(\vx, \vy) \bdel J^b(\vy): ~+~{\cal O} (\vert \vx-\vy\vert )
 \label{sc7d}
 \eeq
The Coulomb potential in (\ref{sc7d}) is of the form
\beq
{\cal V}(\vx, \vy) = m ~{\rm Ein}({\vert \vx-\vy\vert^2 / 2\e}) \approx m \log ({\vert \vx-\vy\vert^2 / 2\e})
\label{sc7e}
\eeq
 where
 \beq
 {\rm Ein} (z) = \int_0^z {dt\over t} (1-e^{-t})\label{sc19a}
 \eeq
In comparing this with  $T {\cal O}_G$, we see that we will get terms similar to $I$ and $II$, but the terms analogous to $III $ and $IV$ will be absent when $\bdel J(\vy)$ is replaced by $\chi(\vy)$ because of (\ref{phi15}). In particular we find
\beqar
I' &=& {e^2 \over 2} \int \Pi_{rs}(\vu,\vv) [\bp_r(\vu), [p_s(\vv), \bdel J^a(\vx)U^{ab}(\vx, \vy)]]
 ~ \chi^b (\vy)~ f(\vx,\vy)\nonumber\\
 & = & \int_{x,y}~ \left[ m + {\cal V}(\vx,\vy)\right]f(\vx,\vy)~\bdel J^a(\vx) U^{ab}(\vx, \vy)  \chi^b(\vy) ~+~{\cal O} (\vert \vx-\vy\vert )
 \label{sc8}
 \eeqar

The normal ordering contribution from the term similar to $II$ is actually zero, due to color contractions.
\beqar
II' &=& {e^2 \over 2} \int \Pi_{rs}(\vu,\vv) \Bigl[ [p_s(\vv), \bdel J^a(\vx)] U^{ab}(\vx, \vy)~ [\bp_r(\vu), \chi^b (\vy)]\nonumber \\
&&\hskip .3in +  \bdel J^a(\vx)[p_s(\vv), U^{ab}(\vx, \vy)]~ [\bp_r(\vu), \chi^b (\vy)] \Bigr] ~f(\vx,\vy)\label{sc7c}
\eeqar
Using the commutation relations (\ref{8}) we find that the first term in $II'$ can be written as
\beqar
II'_{(a)} & = &{e^2 \over 2} \int \Pi_{rs}(\vu,\vv) [p_s(\vv), \bdel J^a(\vx)]~U^{ab}(\vx, \vy)
 ~ [\bp_r(\vu), \chi^b (\vy)]~f(\vx,\vy)\nonumber\\
&&\hskip .1in= m \int_{z,w} \bdel_z \Lambda_{ra} (\vw,\vz) \left[ K(z,\bw ) K^{-1}(w, \bw )\right]_{ab}
f^{brs} \chi^s(\vw) ~f(\vz,\vw)
\label{sc11}
\eeqar
Explicit calculation of $\Lambda_{ra}$ in \cite{KKN1} shows that (see eq.(A3) in Appendix)
\beq
\bdel_z \Lambda (\vw, \vz) =  { e^{-\alpha/2} \over {2  \pi \epsilon}} \Bigl[ K(w, \bw) K^{-1} (z, \bw) +  \sum_{A,B \ge 1} (z-w)^A (\bz-\bw)^B F_{AB} (\vw)~+~ {\cal O}(\epsilon) \Bigr]
\label{sc9}
\eeq
where $\alpha = {{|\vz-\vw|^2} / { \epsilon}}$.
Inserting (\ref{sc9}) into (\ref{sc11}) we find that only the first term in (\ref{sc9}) might contribute a finite $\epsilon^0$-order term, but this is actually zero due to color contraction.
\beqar
II'_{(a)} & = & m \int_{z,w} { e^{-\alpha/2} \over {2 \epsilon \pi}} \left[ K(\vw) K^{-1} (z, \bw)\right]_{ra} \left[ K(z,\bw ) K^{-1}(\vw )\right]_{ab}  f^{brs} \chi^s(\vw) ~f(\vz,\vw) \nonumber \\
& = & 0
\label{sc12}
\eeqar
The second term in (\ref{sc7c}) can be similarly evaluated
\beqar
II'_{(b)} & = &{e^2 \over 2} \int \Pi_{rs}(\vu,\vv) \bdel J^a(\vx)[p_s(\vv), U^{ab}(\vx, \vy)]~ [\bp_r(\vu), \chi^b (\vy)]  ~f(\vx,\vy)\nonumber\\
& = & - { e^2  \over 2} \int \bdel J ^a (\vx) \left[ \Pi _{rs} (\vy; x, \by) - \Pi_{rs} (\vy;  \vy) \right] f^{rbc} f^{sml} K^{al} (x, \by) K^{bm} (y, \by) \chi^c (\vy) f(\vx, \vy)\nonumber\\
\label{sc13}
\eeqar
In writing (\ref{sc13}) we used the fact that
\beq
[p_s (\vv), U^{ab} (\vx, \vy) ] = f^{sml} K^{al} (x, \by) K^{bm} (y, \by) \left[ \delta (v-x) - \delta (v-y) \right] \delta (\bv -\by)
\label{sc14}
\eeq
Explicit calculation shows that
\beq
\Pi _{rs} (\vy; x, \by) - \Pi_{rs} (\vy; \vy) = -{ 1 \over \pi} \sum_{n=1}^{\infty} {{ (x-y)^n} \over {n n!}} \del_y^n K_{rs} (\vy)
\label{sc15}
\eeq
Inserting (\ref{sc15}) into (\ref{sc13}) shows that the contribution of $II'_{(b)}$ is, indeed, negligible for very small separations.
So we find that
  \beq
 (T {\cal O}_G) =  \int_{x,y}~ \left[ m + {\cal V}(\vx,\vy)\right]f(\vx,\vy)~\bdel J^a(\vx) U^{ab}(\vx, \vy) \chi^b(\vy) ~+~{\cal O} (\vert \vx-\vy\vert)
 \label{sc7e}
 \eeq

Finally, the remaining terms in (\ref{sc7a}) can be simplified using
\beq
 [T, V_{YM}] =  2m ~V_{YM}~+~ {{4 \pi} \over { c_A}} \int {\cal{D}} \bdel J {\d \over {\d
J}} \label{sc7f}
\eeq
The term $2m V_{YM}$ is due to both $\bp$ and $p$ acting on $V_{YM}$; this is not needed
for the terms under consideration. We then find
\beqar
&&-{e^2 \over 4m} \int \Pi_{rs}(u,v) \left\{ [\bp_r (u), V_{YM}] [p_s(v), {\cal O}_G]
+[p_s(v), V_{YM}] [\bp_r(u), {\cal O}_G]\right\}\nonumber\\
&& = \int_{x,y}~ \left[ -{\nabla^2_x \over 2m}\right]f(\vx,\vy)~:\bdel J^a(\vx) U^{ab}(\vx, \vy) \chi^b(\vy): ~+~{\cal O} (\vert \vx-\vy\vert )
\label{sc7g}
\eeqar
The $\e \rightarrow 0$ limit of (\ref{phi9}), which has been used to simplify the
right hand side of
(\ref{sc7f}), is adequate for the calculation
of the right hand side of (\ref{sc7g}). Notice that the resulting
factor of $-\nabla^2/2m$, along with
the mass $m$ in (\ref{sc7e}), will reproduce the first two terms of the relativistic combination
$\sqrt{m^2 -\nabla^2}$. The remainder is of order $\vert \vx-\vy\vert$.

Collecting various terms and results, we get
\beqar
\H ~\Psi_G &=& \int_{x,y} ~ \left\{ \left[ M + m -{\nabla_x^2 \over 2m}
+{\cal V}(\vx,\vy)\right] f(\vx,\vy)\right\} ~\bdel J^a(\vx) U^{ab}(\vx, \vy) \chi^b(\vy) ~\Psi_0\nonumber\\
&&\hskip 2.5in ~+~{\cal O} (\vert \vx-\vy\vert )
 \label{sc19}
 \eeqar

 So far, we have not considered the addition of ${\cal Q}$ as in (\ref{phi12}).
 As mentioned earlier, the Coulomb potential in (\ref{sc7e}) is of the form
 ${\rm Ein}({\vert \vx-\vy\vert^2 / 2\e}) \approx \log ({\vert \vx-\vy\vert^2 / 2\e})$.
From the definition of ${\cal Q}$,
\beqar
&&{\cal Q}  \int \bdel J^a(\vx) U^{ab}(\vx, \vy)\chi^b(\vy) ~f(\vx,\vy)\nonumber\\
&&\hskip .8in
 =\e \int \sigma (\vu,\vv, \e)K_{rs}(u, \bv ) [\bp_r (\vu) , \bdel J^a(\vx)] [ p_s (\vv), U^{ab}(x,y)] \chi^b (\vy) f(\vx,\vy)
\nonumber\\
&&\hskip .8in = {c_A \over \pi} \int \bdel J^a(\vx) U^{ab}(x,y)\chi^b(\vy) ~f(\vx,\vy)
\label{sc19b}
\eeqar
where, in the first step, we have indicated the term which gives a nonzero result
among the various terms generated by the action of ${\cal Q}$.
We have evaluated the action of ${\cal Q}$ on $\bdel J^a (\vx) U^{ab}(\vx, \vy) \bdel J^b (\vy)$
before \cite{KKN1}; the nonzero term has the same structure as in (\ref{sc19b}),
except for the replacement of $\chi^b$ by $\bdel J^b$.
Thus, the result in (\ref{sc19b}) is exactly of the same form as obtained
for $\bdel J^a (\vx) U^{ab}(\vx, \vy) \bdel J^b (\vy)$.

The addition of $(e^2/2) \log (2\e /\lambda ) {\cal Q}$ to
the definition of $T$ thus leads to the result
\beqar
\H ~\Psi_G &=& \int_{x,y} ~ \left\{ \left[ M + m -{\nabla_x^2 \over 2m}
+{\cal V}(\vx,\vy, \lambda )\right] f(\vx,\vy)\right\} ~\bdel J^a(\vx) U^{ab}(\vx, \vy) \chi^b(\vy) ~\Psi_0\nonumber\\
&&\hskip 2.5in ~+~{\cal O} (\vert \vx-\vy\vert )
 \label{sc19c}
 \eeqar
 where ${\cal V}(\vx,\vy, \lambda )= m \left( {\rm Ein} ({\vert \vx-\vy\vert^2 / 2\e}) +  \log (2\e /\lambda )\right)
 \approx  m \log ({\vert \vx-\vy\vert^2 / \lambda})$. (Notice that if
 $f(\vx,\vy) $ is of the form $\sigma (\vx,\vy, \lambda )$, the Coulomb potential vanishes.)
Thus, within the approximations used for this continuum strong coupling expansion,
we see that the state $\Psi_G$ becomes an eigenstate of the Hamiltonian ${\cal H}$ with eigenvalue $E$ if $f(\vx,\vy)$ obeys the eigenvalue equation
\beq
\left[ M~+~m - {\nabla_x^2 \over 2m} + {\cal V}(\vx,\vy, \lambda) \right] ~f(\vx,\vy) = E~f(\vx,\vy)
\label{sc20}
\eeq

This equation shows that there is a constituent picture for the gluelump. The gauge-invariant gluon of mass $m$ binds to the external charge via the Coulomb potential.
Since we have arrived at this equation keeping only the $F^2$-term in (Yang-Mills part of)
$\log \Psi_0$, this is appropriate for spatial momenta small compared to $m$. The kinetic energy for the ``gluon'' is thus the nonrelativistic expression. As mentioned before,
 this is the beginning of  a series which sums up to $\sqrt{m^2 - \nabla^2}$
\cite{KKN2}. Further, we find only the Coulomb potential. The key question is whether these approximations are adequate.
If we are interested in a prediction in the continuum strong coupling limit, it is adequate to keep the nonrelativistic kinetic energy term. We cannot say, {\it a priori}, whether this is sufficient for comparison with lattice simulations, since a direct correspondence of the kinematic regimes in the two approaches is not obtained. However, experience with the comparison of the string tension and glueball masses suggests that this is a good starting point.

The question of the potential energy ${\cal V}(\vx,\vy)$  is another
matter. From the action of $T_{YM}$ on $\bdel J (\vx) U(\vx, \vy) \chi
(\vy)$, we get the Coulomb limit of the potential. When effects due
to the vacuum wave function are included, this potential must tend
to the linear potential at large distances. We know this is so, from
the existence of  a nonzero string tension. However, a direct
calculation is difficult and we shall present a somewhat indirect
argument in the next section. We will also use a slightly different 
definition of $\Psi_G$  which is more convenient. 
On the lattice side, the potential
energy is seen to rise linearly with distance, with a value of the
slope or string tension which is in agreement with our calculations,
up to a point, and then it flattens out, indicative of the breaking
of the string. Thus, just at the point where the string breaks, the
constituents of the singlet state (the external charge and the
``gluon'') are separated far enough to be in the regime of the
linear potential. For a sensible comparison with lattice data, we
must understand and incorporate the linear potential. 
We take up this task in the next section. 

However, already at this stage, it is worth noting
that, if we consider the strong binding limit (when the separation
$\vert \vx-\vy\vert$ is negligible) and also the strong coupling limit where
$-\nabla^2 \ll m^2$, the energy
in (\ref{sc20}) can be approximated by $M+m$. The resulting value
$2 (E_0 -M)= 2m$ may be taken as
a lower bound on the string-breaking energy.

\section{The argument for the linear potential}

The strategy we will consider for extracting the linear potential is as follows.
We consider a state of the form
\beq
\vert {\tilde \a} \ra = \int_{x,y} f(\vx,\vy) ~\psi^{\dagger a}(\vx) W^{ab}(\vx,\vy) \phi^{\dagger b}(\vy) ~\vert \a \ra
\label{linpot1}
\eeq
where $W(\vx, \vy)$ is the open Wilson line connecting points $\vx$ and $\vy$, $W(\vx, \vy) = {\cal P} \exp \left(-\int ^{x}_{y} A\right)$ and
$\psi$ and $\phi$ are heavy fields. $\vert \a \ra$ specifies the state of the gauge field
and can be expressed as $\vert \a \ra = {\cal O}_\alpha \vert 0\ra$, where ${\cal O}_\a$
is constructed from the gauge fields only.

The matrix element of $e^{-{\cal H} \tau}$ for states of the
form (\ref{linpot1}) can be related to the expectation value of the Wilson loop; the latter can be expressed in terms of the area law and the linear potential extracted from it. We use two dissimilar fields $\psi$ and $\phi$ to avoid the possibility of mutual annihilation during the (Euclidean) time-evolution.

The action for the theory is taken as
\beq
S = \int d^3x \left[  ({\overline {D\psi}})^a (D\psi )^a +  ({\overline{D\phi}})^a (D\phi )^a - M^2 ({\bar\psi}^a \psi^a
+ {\bar\phi}^a \phi^a ) \right] ~+~ S_{YM}
\label{linpot2}
\eeq
The Hamiltonian has the form
\beq
\H = \H_{YM} + \H_{scalar} +\H_{int}
\label{linpot3}
\eeq
where the last term involves interactions between $\psi$ and $A$ and between
$\phi$ and $A$.
This Hamiltonian commutes with $N_\psi$, the number of $\psi$-particles and
$N_\phi$, the number of
$\phi$-particles, separately. Thus matrix elements of the Hamiltonian are block-diagonal
and we can restrict attention to the $N_\psi = N_\phi =1$ block consistently.
(A real adjoint scalar particle can, in principle, annihilate with another such particle.
This would be an added complication which is irrelevant for
our approach to the linear potential. So we avoid it by choosing
 two distinct complex scalars
to construct the state. This is not the minimal way of doing this.
A minimal way would involve one complex scalar, but the final results are the same,
so we use the more convenient choice of two complex scalars.)

First consider the normalization of states of the form (\ref{linpot1}).
For this, the states are at equal time.
We can use $\psi (\vx) \psi^\dagger (\vy) \vert \a \ra =
\delta (\vx-\vy) \vert \a \ra$, since ${\cal O}_\a$ has no $\psi$'s in it. We then find
\beq
\la {\tilde \a} \vert {\tilde \a} \ra = \int_{x,y} f^*(\vx,\vy) f(\vx,\vy)~ \la \a \vert \a \ra
\label{linpot4}
\eeq
The evaluation of $\la\a \vert \a \ra$ can also be simplified if we approximate the vacuum wave function
by the expression (\ref{sc7}). We then find
\beq
\la \a \vert \a \ra =  \int d\mu ({\cal A} /{\cal G}_*) ~ \exp\left[ {-{1\over 4g^2}\int d^2x F^2}\right]~~{\bar {\cal O}_\alpha} {\cal O}_\alpha
\label{linpot5}
\eeq
We can choose the gauge $A_1=0$, whereupon the action is quadratic in $A_2$ with
a propagator
\beq
\la A^a_2 (\vx) A^b_2 (\vy) \ra = \delta^{ab} \delta (x_1- y_1) \int {dk_2 \over 2\pi}
e^{ik_2(x_2 - y_2)}  ~{1\over k_2^2}
\label{linpot6}
\eeq
The actual evaluation of (\ref{linpot5}) can then be done by Wick contractions using
(\ref{linpot6}).

We now turn to the matrix element $\la {\tilde \b} \vert e^{-\H \tau } \vert {\tilde \a} \ra$, where $\vert {\tilde \b}\ra$ is another state similar to $\vert {\tilde \a}\ra$, with ${\cal O}_\b$ and
a function $h(\vx, \vy)$ replacing ${\cal O}_\a$ and $f(\vx, \vy)$, respectively.
\beqar
\la {\tilde \b} \vert e^{-\H \tau } \vert {\tilde \a} \ra
&=& \int h^*(\vx', \vy') f(\vx,\vy)~\la \b \vert \phi(\vy') W (\vy', \vx')  \psi (\vx')  ~ e^{-\H \tau}~ \psi^\dagger (x) W(\vx,\vy) \phi^\dagger
(y) \vert \a \ra
\nonumber\\
&=&\int h^*(\vx', \vy') f(\vx,\vy)~ \la 0\vert \bigl[{\bar{\cal O}_\b} \phi(\vy') W (\vy', \vx')  \psi (\vx') \bigr]_\tau
 \psi^\dagger (\vx) W(\vx,\vy) \phi^\dagger (\vy)  {\cal O}_\a \vert 0 \ra\nonumber\\
 \label{linpot7}
\eeqar
The set of fields at the left end have been shifted by $\tau$ in Euclidean time.
Since the fields are heavy, we can write, for large $M$, $\phi (\tau) = e^{\H\tau} \phi   e^{-\H \tau}  \approx e^{-M\tau } \phi (0)$, so that
$\la \phi (\tau, \vec{x}') \phi^\dagger (0,\vx)\ra \approx e^{-M\tau } \delta^{(2)}(\vx-\vx')$. We can then simplify the matrix element as
\beq
\la {\tilde \b} \vert e^{-\H \tau } \vert {\tilde \a} \ra
= \int h^*(\vx, \vy ) f(\vx,\vy)~ e^{-2M\tau}~\la 0\vert ~{\bar{\cal O}_\b}(\tau ) ~ W(C) ~ {\cal O}_\a (0) \vert 0\ra
\label{linpot8}
\eeq
where ${\cal O}_\a (0) = {\cal O}_\a (\tau =0)$ and $W(C)$ is the Wilson loop $W(C) = \Tr  {\cal P} \exp (-\oint_C A) $. The path $C$
is a rectangle connecting the points $(\vx , 0)$, $ (\vy , 0)$, $(\vy , \tau )$ and
$( \vx, \tau )$. The evaluation of the remaining expectation value in (\ref{linpot8})
can be understood using Wick contractions.
There are two types of contractions involved: \\
1) Wick contractions within $W(C)$
and separately within ${\bar{\cal O}_\b} {\cal O}_\a$\\
  2) Wick contractions between $W(C)$
and ${\bar{\cal O}_\b} {\cal O}_\a$.\\
We neglect the second type of contractions; we comment on this later.
In this approximation, we have
\beqar
\la 0\vert ~ \bigl[{\bar{\cal O}_\b}\bigr]_\tau ~ W(C) ~ {\cal O}_\a (0)
\vert 0\ra
&\approx& \la 0\vert {\bar{\cal O}_\b}(\tau )  ~{\cal O}_\a (0)\vert 0\ra ~
\la 0\vert W(C)\vert 0\ra\nonumber\\
&=& e^{-\sigma \vert \vx-\vy\vert \tau} ~ \la 0\vert {\bar{\cal O}_\b}(\tau )  ~{\cal O}_\a (0)
\vert 0\ra \nonumber\\
&=& e^{-\sigma \vert \vx-\vy\vert \tau} ~ \la \b \vert ~e^{-\H_{YM} \tau } \vert \a \ra
\label{linpot9}
\eeqar
Once we approximate as in the first line of this equation, the expectation value
of $W(C)$ can be evaluated,
by Euclidean invariance, as an equal-time average.
Also, notice that only $e^{-\H_{YM} \tau}$ appears in the last line.
Thus, within the approximations we have made,
\beq
\la \b \vert \int h^* \phi W \psi ~~ e^{-\H \tau } ~ \int f \psi^\dagger W \phi^\dagger \vert \a \ra
= \int h^* f ~ e^{-2M\tau - \sigma \vert \vx -\vy\vert \tau} \la \b \vert e^{-\H_{YM} \tau }
\vert \a \ra
\label{linpot10}
\eeq
We may rewrite part of this integral as
\beq
\int_{x,y} h^*(\vx,\vy) f(\vx,\vy) e^{-\sigma \vert \vx-\vy\vert \tau}
=\la 0\vert \bigl[\int h^* \phi W \psi \bigr]_{\tau =0}~ e^{-\sigma \vert \vx-\vy\vert \tau}
\bigl[ \int f \psi^\dagger W \phi^\dagger\bigr]_{\tau =0} \vert 0 \ra
\label{linpot11}
\eeq
Using this result, we finally get
\beq
\la \b \vert \int h^*  \phi W \psi~~ e^{-\H\tau} \int f \psi^\dagger W \phi^\dagger \vert \a \ra
= \la {\tilde \b}\vert  e^{- (2M +\sigma \vert \vx-\vy\vert )\tau} \int f \psi^\dagger W \phi^\dagger
e^{-\H_{YM} \tau} \vert \a \ra
\label{linpot12}
\eeq
Since $\vert {\tilde \b}\ra$ is completely arbitrary within the $N_\psi = N_\phi =1$ block,
we conclude, by taking terms linear in $\tau$,
\beq
\H ~\int f \psi^\dagger W \phi^\dagger \vert \a \ra =
\left( 2 M  + \sigma \vert \vx -\vy\vert \right) ~ \int f \psi^\dagger W \phi^\dagger \vert \a \ra  ~+ \int f \psi^\dagger W \phi^\dagger ~ \H_{YM} {\cal O}_\a \vert 0\ra
\label{linpot13}
\eeq

Some comments on the approximations used in this calculation are in order at this point.
\begin{enumerate}
\item We have approximated the vacuum wave function by equation (\ref{sc7}).
This is adequate for the linear potential, but there can be corrections due to the higher
terms (in powers of momenta and powers of $J$) in the exponent.
We hope to address the issue of higher order terms elsewhere.
\item Wick contractions between $W(C)$ and ${\bar{\cal O}_\b} {\cal O}_\a$ have been neglected.
These can lead to structures different from $W(C)$  on the right hand side and hence to nondiagonal pieces for the first term in (\ref{linpot13}). These are small if the linear potential dominates. The lattice data suggest that this is the case, as explained earlier.
Further, since $W(C)$ has no color charges, these contractions will be suppressed by
powers of $1/N$.
\item We need large $M$ limit to avoid virtual and real processes with additional
numbers of $\psi$ and $\phi$ particles. Such processes could also modify the Yang-Mills vacuum, if we do not take the large $M$ limit. (This is a quenched approximation
for these particles.)
\end{enumerate}

Consider now the case of the gluelump. To get the wave function of the gluelump,
we can take equation (\ref{linpot13})  with ${\cal O}_\a =1$ and apply on this equation
the operator $\int \bdel J^a \eta^a$, where
$\eta^a = M^{\dagger ak}\psi^k $. This has the effect of replacing
$\psi^\dagger$ by $\bdel J^a M^{\dagger ak}$. Further the action of
$\phi^\dagger$ on the vacuum wave function for the heavy scalars
gives $\phi = M^{\dagger -1} \chi$. Thus we find
\beqar
\int \bdel J \eta  ~\psi^\dagger W \phi^\dagger \Psi_0 
&=& \bdel J^a (\vx) \left( M^{\dagger ak} (\vx) W^{kl}(\vx,\vy) M^{\dagger -1 lb}(\vy) \right)\chi^b(\vy)
\Psi_0\nonumber\\
&\equiv& \bdel J^a (\vx) ~{\widetilde W}^{ab} (\vx,\vy)~ \chi^b (\vy) \Psi_0
\label{linpot13a}
\eeqar
The operator ${\widetilde W}^{ab} (\vx,\vy)$ may be written as
\beq
{\widetilde W}^{ab} (\vx,\vy) = \left[ {\cal P} \exp \left( \int_y^x \del K K^{-1}\right)\right]^{ab}
\label{linpot13b}
\eeq
It is an open Wilson line for the holomorphic component of the connection $\del K K^{-1}$.
(The result (\ref{linpot13b}) follows from the fact that we may 
think of $(A, \bA)$ as a complex gauge transform (by the matrix $M^\dagger$)
of $(-\del H H^{-1}, 0)$, i.e., $(A, \bA) = (-\del H H^{-1} , 0)^{M^\dagger}$, and
because $W^{kl}(\vx,\vy, A^g) = g^{-1}(\vx) W^{kl}(\vx,\vy) g(\vy)$ for
$A^g = g^{-1} A g + g^{-1} \del g$.)

As in the case of $W$, ${\widetilde W}$ depends on the curve chosen from $\vy$ to $\vx$.
For a short separation $\vert \vx - \vy \vert$, we can write
\beq
{\widetilde W}(\vx,\vy) = K(x, \by ) K^{-1} (y, \by ) =  U(\vx,\vy)
\label{linpot13c}
\eeq
so that the functional $\bdel J^a {\widetilde W}^{ab} \chi^b$ agrees with the gluelump
state used in section 4. For a longer separation, there are differences. Using the composition property of path-ordered exponentials, we find, for a straight-line path,
\beqar
{\widetilde W}(y+2 \e ,y) &=& K(y +2 \e , \by + {\bar \e}) ~K^{-1}(y+\e, \by + {\bar \e})
~K(y+\e, \by ) ~K^{-1} (y, \by ) \nonumber\\
&\approx& K(y+2\e, \by ) ~K^{-1}(y, \by ) + \e {\bar \e}~ \bdel (\del K K^{-1}) +\cdots\nonumber\\
&=&U(y+2\e , \by ) + \e {\bar \e} ~\bdel (\del K K^{-1}) +\cdots
\label{linpot13d}
\eeqar
Thus the Wilson line ${\widetilde W}$ will differ from $U(\vx,\vy) = K(x, \by ) K^{-1}(y, \by)$
by terms involving $\bdel (\del K K^{-1})$ and higher derivatives of it when the series is continued. Such terms evidently correspond to deformations of the path.

The gluelump state can be defined, at finite separations, by either 
$\bdel J^a (\vx) U^{ab} (\vx,\vy) \chi^b (\vy)$ or
$\bdel J^a (\vx) {\widetilde W}^{ab} (\vx,\vy) \chi^b (\vy)$. Which is more advantageous
will depend on which leads to a simpler equation within the approximations we are using.
The emergence of equation (\ref{linpot13}) suggests that 
$\bdel J^a (\vx) {\widetilde W}^{ab} (\vx,\vy) \chi^b (\vy)$ is a simpler functional to use,
and so we redefine the gluelump as
\beq
\Psi_G = \int d^2xd^2y~ f(\vx,\vy)~\bdel J^a(\vx)~  {\widetilde W}^{ab} (\vx,\vy)~ \chi^b(\vy) ~~\Psi_0
\label{linpot13e}
\eeq
Using (\ref{linpot13a}), we then find
\beq
\H ~\int f \bdel J {\widetilde W} \chi ~\Psi_0 =
[ \H, \int \bdel J \eta ] \int f \psi^\dagger W \phi^\dagger \Psi_0
+
\left(2 M  + \sigma \vert \vx -\vy\vert \right) ~ \int f \bdel J {\widetilde W} \chi ~\Psi_0
\label{linpot14}
\eeq
The operator $\bdel J \eta$ is a local operator. This must be defined,
in a regularized way, by
an expression of the form ${\cal O}_G$ where $f(\vx,\vy) $ is replaced by
$\sigma (\vx, \vy, \lambda )$; then we must take the small $\lambda$ limit.
In this case, we can use the result (\ref{sc19c}); the Coulomb potential is zero
and we can write
\beq
[ \H, \int \bdel J \eta ] \Psi_0
= \int \sigma (\vu, \vv, \lambda ) \left[ \left(-M + m  - {\nabla_u^2 \over 2m}\right)
\bdel J^a(\vu)\right] U^{ab}(\vu, \vv) \eta^b (\vv) \Psi_0~+ \cdots
\label{linpot15}
\eeq
where $U^{ab}(\vu, \vv)$ is used because of the point-splitting needed to 
define $\bdel J \eta$.
The omitted terms in (\ref{linpot15}) correspond to terms with either $\bp$ or $p$
acting on $\int \bdel J \eta$ and the other left free so that it can act on $\int f \psi^\dagger W \phi^\dagger $ in (\ref{linpot14}); i.e., they are of the form
\beqar
&&{e^2 \over 2} \int \Pi_{rs}(u,v)\left\{  [ \bp_r (u), \int \bdel J \eta ]
~ f \psi^\dagger [p_s (v), W ] \phi^\dagger
+ [p_s(v), \int \bdel J\eta ]~ f \psi^\dagger [\bp_r(u), W ] \phi^\dagger 
\right\} \Psi_0\nonumber\\
&&= {e^2 \over 2} \int \Pi_{rs}(u,v)\left\{  [ \bp_r (u), \int \bdel J \eta ]
~f \eta^\dagger [p_s (v), {\widetilde W} ] \chi^\dagger
+ [p_s(v), \int \bdel J\eta ]~ f \psi^\dagger [\bp_r(u), W ] \phi^\dagger 
\right\} \Psi_0\nonumber\\
\label{linpot15a}
\eeqar
One of the terms on the right hand side, namely, ${e^2 /2} \int \Pi_{rs}(\vu,\vv) [ \bp_r (\vu), \int \bdel J^a ]  [p_s (\vv), {\widetilde W}^{ab} ] \chi^{\dagger b}$ is the type of term which led to the
Coulomb potential in the calculation of section 4.  So one might wonder whether this can generate a Coulomb term in addition to the linear potential.
This is not the case. In fact, the Coulomb potential is due to the term
$f_{arm} \bdel J^m \delta (\vx-\vu)$ in the commutator
$[\bp_r (\vu) , \bdel J^a (\vx) ] = -i \bdel \D_{ar} \delta (\vx-\vu)$. Notice that in
(\ref{linpot15a}) we have
the combination $[\bp_r (\vu), \int \bdel J \eta ]$ and, since $[ \bp_r (\vu), \eta^a (\vx)]
=f_{arm} \eta^m \delta (\vx-\vu)$ as well, the relevant term is actually zero.
(This too has to be understood with point-splitting, but this particular result is 
basically the same as in the unregulated calculation.)
We will neglect the remaining terms in (\ref{linpot15a}), which lead to new 
operator structures.
These terms lead to operator structures with more powers of $J$. The gluelump state we consider can have overlap with states which have the same quantum numbers as $\bdel J \widetilde{W} \chi$. The fact that the Hamiltonian acting on $\bdel J \widetilde{W} \chi$ can evolve it into other structures is indicative of this possibility of mixing. However, as is well known in quantum mechanics, the effect of such nondiagonal terms in ${\cal H}$ comes with energy denominators and can be taken to be small if the differences between the energies of the lowest gluelump state (which is what we are interested in) and higher states are large enough. We expect this to be the case at large enough coupling, since the differences must go like $m$. But 
ultimately it is to be justified {\it a posteriori}.\footnote{The transition amplitude from the unbroken string state to the gluelumps is seen to be very small from lattice data, even for $SU(2)$ \cite{PW, FK}. This is another indication that the off-diagonal elements mentioned above might be small.}

With this understanding of various terms, combining 
the results in (\ref{linpot14}) and (\ref{linpot15}), we finally have, as $\lambda$ in (\ref{linpot15}) is
taken to be very small,
\beq
\H ~\Psi_G  =
\int \left[ M+ m  - {\nabla_x^2 \over 2m} + \sigma \vert \vx -\vy\vert
\right] f(\vx,\vy)~ \bdel J^a (\vx) {\widetilde W}^{ab}(\vx, \vy)  \chi^b (\vy)
\Psi_0
~+ \cdots
\label{linpot16}
\eeq
The approximations involved in this equation are given by the comments after
equation (\ref{linpot13}) and by the comments
about the omitted terms in (\ref{linpot15}, \ref{linpot15a}).

Equation (\ref{linpot16}) shows that the eigenvalue equation we need for the gluelump
states is
\beq
\left[ M+ m  - {\nabla_x^2 \over 2m} + \sigma \vert \vx -\vy\vert
\right] f(\vx,\vy) = E ~f(\vx,\vy)
\label{linpot17}
\eeq
By our general arguments, the string-breaking energy is given by
$V_* = 2 (E_0 -M )$ where $E_0$ is the ground state energy eigenvalue of the equation
(\ref{linpot17}).
We now turn to the determination of this value.

\section{A variational estimate for string breaking}

Since we are interested in the ground state eigenvalue, the simplest
estimate is given by a variational calculation.
Removing the center of mass motion (which is zero as $M \rightarrow \infty$)
the energy functional corresponding to (\ref{linpot17}) is
\beqar
E &=&  M + m + {1\over {\cal N}} \int d^2x~ \left[ {\vert \nabla f \vert^2 \over 2m}  ~+ \sigma
\vert \vx\vert ~\vert f\vert^2 \right] \nonumber\\
{\cal N} &=& \int d^2x~ \vert f \vert^2 \label{var1} 
\eeqar 
We take
a variational  ansatz of the form $f = \exp ( - \beta \vert x\vert
^{\mu} )$, where both $\beta $ and $\mu $ are regarded as
variational parameters.  The minimum of $E(\beta , \mu)$ with respect to $\beta$
is found to be at $\beta = \beta_*$,
\beq
\beta_* = \left( \frac{2^{(\mu -3)/\mu}(2 m
\sigma )\Gamma(3/\mu )}{\mu ^2}\right)^{\mu/3}
\label{var1a}
\eeq
A formula for the
energy eigenvalue $E(\beta_*, \mu)$ as a function of $\mu $ can now be obtained.
\begin{eqnarray} 
E(\beta_*, \mu) &=& M + m +\Biggl[  \frac{2^{-(2\mu +1)/\mu}}{2m\Gamma (2/\mu)}\left(\frac{2^{(\mu
-3)/\mu}2m\sigma \Gamma(3/\mu)}{\mu ^2}\right)^{-1/3}\nonumber\\
&&\hskip .2in\times \left(8m\sigma
\Gamma(3/\mu ) + 8^{1/\mu }\mu^2 \left(\frac{2^{(\mu
-3)/\mu}2m\sigma \Gamma(3/\mu)}{\mu^2}\right)\right)\Biggr]
\label{var2}
\end{eqnarray} 
The string tension for a representation $R$ is given by (\ref{intro2});
for the present
calculation, we need the adjoint string tension, $\sigma_A = \pi
m^2$. Using this value of the string tension, the formula for $E(\beta_*, \mu)$
can be minimized numerically. The minimum is obtained at $\mu_* =
1.752$. This gives $E_0 -M = E(\beta_*, \mu_*) -M =   3.958m$, corresponding to a 
string-breaking energy of $V_{*cal} = 2(E(\beta_*, \mu_*) -M ) = 7.916~m$.

To check the robustness of the variational result, we have also
tried to estimate the string breaking energy using a more general
variational ansatz of the form 
\beq
f = \frac{e^{-\beta \vert x
\vert^\mu}}{(1+ \vert x\vert )^\nu }
\label{var2a}
\eeq
Such an ansatz is suggested
by the large $x$ solution of the two-dimensional Schr\"odinger equation
with a linear potential, whose solutions fall off, at large $\vert x\vert$, as
\beq
\frac{{\rm Ai}\left((2m \sigma)^{1/3}\vert x \vert\right)}{\sqrt{\vert x \vert
}}\approx \frac{\exp \left[ {-\frac{2}{3} \left((2m\sigma)^{1/3}\vert x
\vert\right)^{3/2}}\right]}{\vert x \vert ^{3/4}} \label{var2b}
\eeq
where ${\rm Ai}$ is the Airy function. However, a
numerical minimization of $E(\beta , \mu, \nu)$, with respect to the three variational
parameters $\beta, \mu, \nu$, shows that the lowest estimate for
$E$ is provided by  $\nu = 0$, i.e., by the exponential ansatz reported
above.

We are now in a position to compare this with lattice estimates. The
latest data on this is for $SU(2)$, given by Kratochvila and Forcrand
\cite{FK}. The string-breaking energy is given by $V_{*lat}
\approx 8.68 ~m$. This arises as follows.
$V_{*lat}$ is computed in terms of the lattice spacing $a$ as
$V_{*lat} \approx 2.063 a^{-1}$. In the same set-up, the
fundamental string tension is given by $\sigma_F = 0.0625 a^{-2}$.
This can be used to write $V_{*lat} \approx 2. 06 ~m~ \sqrt{ \pi c_F /
0.0625 c_A} \approx 8.68~m$. For our analytic calculation, we then have
$\vert V_{*lat} - V_{*cal}\vert / V_{*lat} \approx 8.76\%$.

It is interesting also to make a similar estimate for the $0^{++}$
glueball mass, at least as a further check on the approximations used. 
If we use our picture of two constituents of mass $m$
each, interacting via the linear potential, the eigenvalue equation
is 
\beq 
\left[ 2 m  - {\nabla_x^2 \over 2m} - {\nabla_y^2 \over 2m}+
\sigma \vert \vx -\vy\vert \right] f(\vx,\vy) = E ~f(\vx,\vy) \label{var4} 
\eeq
The only difference, compared to what we have done for the gluelump, is that the
reduced mass should now be $\half m$. This leads to 
\beq 
M_{0^{++}} = 5.73~m\label{var5} 
\eeq 
The difference with the lattice estimate of $5.17
~m$ \cite{teper} is  $10.83\%$. In this case, of course, since both
constituents have mass $m \ll M$, our $1/m$-expansion and the
nonrelativistic approximation are less reliable.

\section{Discussion}

A Hamiltonian approach to Yang-Mills theory in 2+1 dimensions was introduced several years ago by two of the authors \cite{KN, KKN1,KKN2}. In this paper we have extended the formalism to include matter fields, specifically scalar fields in the adjoint representation. First of all, this may be viewed as a first step
towards analyzing supersymmetric theories.  
Secondly, it may be used for analyzing screening and string breaking.
On general group-theoretic grounds we should expect screening of Wilson 
loops in the adjoint and other representations of zero $N$-ality. 
When the length of a string in one of the screenable representations is increased, it breaks
at some point with the formation of two color singlet states, each of which is
referred to as a gluelump. We obtained a Schr\"odinger equation for a gluelump
state, the variational solution of which then gave an analytic estimate
of the string-breaking energy. This is within $8.8\%$ of the latest lattice estimates. The emerging picture is that of a bound state between a constituent ``gluon" $\bdel J$, of mass $m$, and a heavy adjoint scalar of mass $M$, interacting via a linear potential $V \sim \sigma_{A} r$.
Some approximations are necessary for obtaining the Schr\"odinger equation,
since most of the states under discussion are unstable; the approximations we have used
are given in section 5. It should also be emphasized that we are calculating the energy at which the string breaks, the
transition amplitude for the breaking is not calculated. (These are two different questions.
The transition amplitude, as is well known, will be
suppressed at large $N$.)

Specifically for the $(2+1)$-dimensional theory, the
string-breaking energy has not been analyzed extensively on the lattice except for
the group $SU(2)$.
We consider this to be an important question for which more lattice data are needed,
for higher groups, and, in particular, the $N$-dependence.

Also, as explained in the text, string breaking is closely related to the issue of corrections
to the formula (\ref{intro2}) for the string tension. This matter will be taken up in a future
publication.
\vskip .1in
VPN thanks David Gross for discussions at a preliminary stage of this
work during a visit to KITP, Santa Barbara in 2004.
We also acknowledge useful discussions with Barak Bringoltz.
AA also wishes to thank the Max Planck Institut fur
Gravitationsphysik, where part of this work was carried out, for hospitality.
This research was supported in part by the National Science Foundation 
grants PHY-0457304 and PHY-0555620 and by PSC-CUNY grants.
\section*{\normalsize APPENDIX}
\def\theequation{A\arabic{equation}}
\setcounter{equation}{0}
Once the regulated Green's functions (\ref{phi7}) have been introduced, it is straightforward to evaluate the regularized form of the Hamiltonian and its action on various
functionals.
However, it is important to understand and follow the correct order of limits.
Here we shall comment briefly on this issue.

We start by recalling some elementary properties of a covariant integral $I(p, \e )$ resulting from some term in the perturbation expansion of a field theory. Here $p$ denotes external momenta and $\e$ is the short-distance cut-off.
Once the integration over the loop momenta are done, $I(p,\e )$ is of the form
\beq
I(p, \e )= \sum_\a \e^{-\a} ~{\cal O}_\alpha (p) ~+~ {\tilde {\cal O}}(p)
~+~ \sum_\beta \e^\beta {\cal O}'_\beta
\label{A1}
\eeq
The first set of terms involves negative powers of $\e$ and represent potentially divergent
terms as $\e$ becomes very small.
(We include $\a =0$, which may be taken as
a logarithmic divergence.) ${\tilde {\cal O}}(p)$ represents finite terms and
the last set of terms vanish as $\e \rightarrow 0$. At finite, but small, $\e$, the
last set of terms are suppressed by powers of $p^2 \e$; they are negligible for
small external momenta. While the divergent and finite terms are independent of the regulator (apart from the usual ambiguity of the subtraction points), the
${\cal O}'_\beta$ terms are sensitive to the regulator.
Of course, as $\e \rightarrow 0$, this does not matter, but notice that for external momenta
of the order of $\e^{-1/2}$, these terms can contribute and constitute regularization 
ambiguities. The correct procedure is obviously that
the result (\ref{A1}) is to be used only for processes with external momenta
small compared to $\e^{-1/2}$.

A simple case which highlights this point is the following. A regulator does not have to respect symmetries, say, Lorentz invariance; the standard lattice regularization is an example. In a Lorentz-invariant theory, this can lead to
Lorentz-violating terms of order $p\sqrt{\e}$. If we consistently restrict our calculation to processes with momenta $p \ll \e^{-1/2}$, these spurious regulator-dependent terms will
be negligible.

In a Hamiltonian formulation, we define the regulated kinetic energy operator
$T(\e )$, given by (\ref{phi6}), (\ref{phi7}). Since it is a functional differential operator, to interpret it properly, we have to consider its action on different functionals.
Consider a functional $\Psi(\lambda' )$ with the average separation of points
between fields in the functional being $\sqrt{\lambda'}$. When we act on this with
$T(\e )$, there are three types of terms possible.
\begin{enumerate}
\item Terms which diverge as $\e \rightarrow 0$. If there are such terms, we must introduce a subtraction procedure to make $T(\e )$ well-defined and eliminate these terms. (The redefinition of $T(\epsilon)$ with the addition of ${\cal Q}$ in (\ref{phi12}) may be viewed as an example of this.)
\item Terms which are finite as $\e \rightarrow 0$. These are physical effects.
\item Terms which are of order $\e$. These can be neglected. However, if we take
$\lambda' \sim \e$, or $\lambda' \rightarrow 0$ at finite $\e$, these terms can give
finite results, just as the ${\cal O}'_\beta$ terms in (\ref{A1}).
Exactly as in the previous case, the correct procedure is to realize that the $T(\e )$ can only be used for functionals with $\lambda' \gg \e$; otherwise there will be spurious regulator-dependent terms.
\end{enumerate}
Keeping this in mind, all the calculations have to be done with $\e \rightarrow 0$ first.
A natural question is then: How do we define a ``local" operator like $V_{YM}\sim \int B^2$?
Note that this is needed for the Hamiltonian, and further we will need the commutator of
$T$ with this operator. From what was said before, $V_{YM}$ must be defined in a regulated way with an average point-separation of $\sqrt{\lambda}$, with $\lambda \gg \e$.
Then the action of $T(\e )$ on this is correctly obtained. To make the operator
local, we then take $\lambda \rightarrow 0$. In other words, we must take
$\e, \lambda$ to zero, keeping $\lambda \gg \e$. This is the correct procedure
and it will yield unambiguous results.
This procedure is even more crucial in a Hamiltonian framework;
otherwise, spurious terms may appear and it may not even be obvious that they are spurious,
since Lorentz symmetry is not manifest.

This procedure was used in \cite{KKN1, KKN2}. For the purposes of this paper, we note that
an expression for $\Lambda_{ra}$ was obtained in \cite{KKN1} as
\beq
\Lambda _{ra} (\vw,\vz) =  {1 \over {\pi (z-w)}} \bigl[ \d _{ra} - \bigl( K(\vw) K^{-1}
( u, \bw )K (\vu) K^{-1} (z, \bu) \bigr) _{ra} e^{- \a /2} 
\bigr]  + {\widetilde{\Lambda _{ra}} }(\vw,\vz)
\label{A2}
\eeq
\beqar
{\widetilde{\Lambda _{ra}} }(\vw,\vz)&=& {\e \over \pi} e^{- \a /2} \bigl[  -2 \del _z { 1 \over {z-w}}
\bdel _z \bigl( K(\vw) K^{-1} ( u, \bw )K (\vu) K^{-1} (z, \bu) \bigr) \nonumber\\
&& +  { 2
\over {z-w}} \bdel _z \bigl( K(\vw) K^{-1} ( u, \bw )K (\vu) \del _z K^{-1} (z, \bu)
\bigr)   \nonumber\\
&& + { 2 \over {(z-w) (\bz -\bw)}}   K(\vw) K^{-1} (u,\bw) K(\vu) \del _z
\bigl( K^{-1} (z, \bu) K(\vz) \bigr) K^{-1} (\vz)    \bigr]  _{ra} \nonumber\\
&&\hskip .3in + {\cal O} (\e ^2) 
\label{A3}
\eeqar
where $u =\half (z+w),~\bu = \half (\bz + \bw)$ and $\alpha = (z-w) (\bz - \bw)/\epsilon$. 
If we take $\sqrt{\e}$ to be very small compared to the average separation of fields
in functionals on which $T$ acts, ${\widetilde{\Lambda _{ra}} }(\vw,\vz)$ does not contribute
and we can use the simplified form for $\Lambda_{ra}(\vw, \vz )$, namely,
expression (\ref{A2}) with ${\widetilde{\Lambda _{ra}} }(\vw,\vz)$ set to zero. In fact we can simplify the first term even further, approximating $\Lambda_{ra} (\vw, \vz)$ as
\beqar
\Lambda _{ra} (\vw,\vz) & = & {1 \over {\pi (z-w)}} \bigl[ \d _{ra} - \bigl( K(\vw) K^{-1}
( z, \bw )\bigr) _{ra} e^{- \a /2}  \bigr]  \nonumber \\
 &+& ( {\rm
terms~of~higher~order~in~\e~or~(z-w),~(\bz - \bw)}) 
\label{A4}
\eeqar
This is correct and adequate, but it must be kept in mind that $T$ so defined acts on operators with a point-splitting separation $\sqrt{\lambda} \gg \sqrt{\epsilon}$, and in taking the small $\lambda$-limit we must preserve this inequality.

\end{document}